\documentstyle[12pt]{article}
\newcommand{\at}[2]{{\raisebox{-1ex}{\footnotesize{#1}} \atop \raisebox{1ex}{\footnotesize{#2}}}}
\newcommand{\last}{\mbox{\large ${\ast}$}}

\begin{document}
\begin{flushright}
CU/TP-96-5 \\
August 1996 
\end{flushright}
\vspace{0.5cm}
\begin{Large}
\vspace*{1cm}
\centerline{\bf Violation of universal Yukawa coupling}
\centerline{\bf and quark masses}
\end{Large}
\vspace{2cm}
\centerline{Tadayuki Teshima and Toyokazu Sakai }
\vspace{1cm}
\centerline{\it Department of Applied Physics,  Chubu University}
\vspace{0.3cm}
\centerline{\it Kasugai, Aichi 487, Japan}
\vspace{2cm}
\setlength{\baselineskip}{0.85cm}
\centerline{ABSTRACT}
\par 
We analyse the quark mass hierarchy and CKM matrix using the universal Yukawa coupling model with small violations precisely. We estimate the ranges of the values of 8 violation parameters ($\delta_1^u$,\ $\delta_2^u$,\ $\delta_3^u$,\ $\delta_1^d$,\ $\delta_2^d$,\ $\delta_3^d$,\ $\phi_2$,\ $\phi_3$) in our quark mass matrices satisfying quark mass ratios and CKM matrix, where $\phi_2$,\ $\phi_3$ are phases. Without these phases, the solution satisfying quark mass ratios and CKM matrix is not obtained. These parameters obtained can explain the CP violation effects.

\newpage
\setlength{\baselineskip}{0.85cm}

\begin{large}
\par \noindent
{\bf 1. Introduction}	
\end{large}
\vspace{0.5cm}

\par
The origin of the mass hierarchy of quarks and leptons has been investigated by the various theories beyond the standard model (SM) by many authors \cite{FRITZSCH}-\cite{HABA}. Although it is necessary to study precisely the theories beyond the SM, in the present circumstances where there are the precise analysis of $B^0-\bar{B}^0$ mixing , the CP-violating parameter $\varepsilon$ of the $K^0-\bar{K}^0$ system and the determination of the top-quark mass, one should analyse the mass hierarchy of quarks and Cabibbo-Kobayashi-Maskawa (CKM) matrix independently of model assumed. Furthermore, many attempts beyond SM \cite{HABA} constructing mass matrix patterns at the GUT scale in SUSY theories or at the string scale in string models, although quite successful, cannot produce results in complete agreement with precise low energy data. Thus, for the model building beyond SM, the analysis using only the minimal qualities to be sure at present is now to be very necessary. 

\par
For quark mass matrix patterns in low energy, there are Fritzsch type, Stech type model \cite{FRITZSCH}, the  democratic type model \cite{HARARI} and the universal Yukawa coupling type model \cite{GATTO}. We adopt a quark mass matrix like the democratic \cite{HARARI} and the universal Yukawa coupling type models \cite{GATTO} with the small violations from the universality which cause the mass hierarchy. Our model does not take any assumptions on the violations and treats violation parameters as free parameters. First, we see the mass hierarchy mechanism in the limit of the universal Yukawa coupling. The $(u,\ c,\ t)$ and $(d,\ s,\ b)$ quark mass matrices are expressed, under the universality of Yukawa coupling strength, as 
\begin{equation}
  M^q = \Gamma^q \left(
  \begin{array}{ccc}
  1 & 1 & 1 \\
  1 & 1 & 1 \\ 
  1 & 1 & 1
  \end{array} 
  \right), \ \ \ (q=u,\ d)
\end{equation}
where $\Gamma^u$ and $\Gamma^d$ are real, and are not assumed universal. This $\Gamma^u$ and $\Gamma^d$ unuiversality is guaranteed by e.g., a minimal supersymmetric gauge model \cite{NILLES} in which the up and down quarks acquire their masses through the couplings to two different Higgs multiplets. It is well known that this type of the mass matrix is diagonalized as  ${\rm diag}[0,~0,~3\Gamma^{q}]$ by the orthogonal matrix $T_0$; ${\rm diag}[0,~0,~3\Gamma^{q}]=T_0 M^q T_0^{-1}$, where $T_0$ is\begin{equation}
  T_0 = \left(
  \begin{array}{ccc}
  \frac1{\sqrt{2}} & -\frac1{\sqrt{2}} & 0 \\
  \frac1{\sqrt{6}} & \frac1{\sqrt{6}} & -\frac2{\sqrt{6}} \\ 
  \frac1{\sqrt{3}} & \frac1{\sqrt{3}} & \frac1{\sqrt{3}}
  \end{array} 
  \right). 
\end{equation}
Thus the type of Eq.~(1) gives the large mass gap between the heaviest quark and other two quarks. 

\par
Next, we introduce small violations with phases of Yukawa coupling strength in universal coupling (Eq.~(1)) as
\begin{equation}
  M^q = \Gamma^q \left(
  \begin{array}{ccc}
  1 & 1-\delta^q_1 e^{i\varphi^q_1} &  1-\delta^q_2 e^{i\varphi^q_2} \\
  1-\delta^q_1 e^{-i\varphi^q_1} & 1 &  1-\delta^q_3 e^{i\varphi^q_3} \\ 
  1-\delta^q_2 e^{-i\varphi^q_2} & 1-\delta^q_3 e^{-i\varphi^q_3} & 1
  \end{array} 
  \right), \ \ \ (q=u,\ d)
\end{equation}
where $\delta^q_i$ are small real violation parameters  
\begin{equation}
\delta^{u,d}_i \ll 1. \ \ \ \ (i = 1,\ 2,\ 3)  
\end{equation}
We do not take any assumptions on the violation parameters except that $\delta^q_i$'s  are very small. Introducing small violations of coupling, we assumed that the violations are caused from the coupling between different quarks. Here it should be stressed that the large mass differences are produced by the universal coupling then what distinguishes the quarks is not the masses but other characters. Thus the assumption that the diagonal elements of couplings between same quarks are same each other and do not have violations is reasonable. In the parametrization (Eq.~(3)), we used the $-$ sign before $\delta^q_i$ because, in this notation, $\delta^q_i$ are allowed only to be positive in the quark mass analysis as shown later. Branco, Silva-Marcos and Rebelo \cite{GATTO} studied the type of mass matrix (3) but they equate the type of this mass matrix to the quark mass squared. 
     
\par
Under the assumption of small violation in universal coupling, we can get the second mass gap between two degenerate zero mass states which are taken from the universal mass matrix (1). This is shown from the mass matrix (3) neglecting phases 
\begin{equation}
  M^q = \Gamma^q\left(
  \begin{array}{ccc}
  1 & 1-\delta^q_1 & 1-\delta^q_2 \\
  1-\delta^q_1 & 1 & 1-\delta^q_3 \\ 
  1-\delta^q_2 & 1-\delta^q_3 & 1
  \end{array} 
  \right).\ \ \ \ \ (q=u,\ d)  
\end{equation}
This mass matrix is transformed by $T_0$ to 
\begin{eqnarray}
  &&T_0M^qT_0^{-1} =  \nonumber \\
  &&\ \ \  \Gamma^q\left(
  \begin{array}{ccc}
  \delta^q_1 & (\delta^q_2-\delta^q_3)/\sqrt{3} & (-\delta^q_2+\delta^q_3)/\sqrt{6} \\
  (\delta^q_2-\delta^q_3)/\sqrt{3} & (-\delta^q_1+2\delta^q_2+2\delta^q_3)/3 & (-2\delta^q_1+\delta^q_2+\delta^q_3)/{3\sqrt{2}} \\ 
  (-\delta^q_2+\delta^q_3)/\sqrt{6} &  (-2\delta^q_1+\delta^q_2+\delta^q_3)/{3\sqrt{2}} & (9-2\delta^q_1-2\delta^q_2-2\delta^q_3)/3
  \end{array} 
  \right), 
\end{eqnarray}
then if $\delta^q_1 \ll \delta^q_2 \approx \delta^q_3 \ll 1$, three eigenvalues turn to $(0,\ 4\delta^q_2\Gamma^q/3,\ (3-4\delta^q_2/3)\Gamma^q)$, approximately. This tendency is certified and the allowed ranges of $(\delta^q_1,\ \delta^q_2,\ \delta^q_3)$ are determined precisely from the analysis of quark mass ratios and CKM matrix in the following numerical study.
\par
In last section (3. Discussions), we will comment on the difference between the results of our model and others \cite{FRITZSCH}-\cite{GATTO}    
 
\vspace{1.2cm}

\begin{large}
\par \noindent
{\bf 2. Numerical analysis}
\end{large}
\vspace{0.5cm}

\par
 The mass matrices (3) contain 6 violation parameters for $(u,\ c,\ t)$ and $(d,\ s,\ b)$ sector except for the $\Gamma^q$, respectively. First, we consider 3 parameters $(\delta^q_1,\ \delta^q_2,\ \delta^q_3)$ case (Eq.~(5)) for simplicity and later we will consider the case containing phases. We diagonalize the mass matrices (5) to the ${\rm diag}[m_u,\ m_c,\ m_t]$ and  ${\rm diag}[m_d,\ m_s,\ m_b]$ for $q=u$ and $d$ by the unitary matrices $T(\delta^u_1,\ \delta^u_2,\ \delta^u_3)$ and $T(\delta^d_1,\ \delta^d_2,\ \delta^d_3)$, respectively, 
\begin{eqnarray}
&&T(\delta^q_1,\ \delta^q_2,\ \delta^q_3)M^qT^{-1}(\delta^q_1,\ \delta^q_2,\ \delta^q_3)=M^q_D\ ,\ \ \ \ (q=u,\ d) \nonumber \\
&&M^u_D={\rm diag}[m_u,\ m_c,\ m_t], \ \ \ \ \ M^d_D={\rm diag}[m_d,\ m_s,\ m_b]. 
\end{eqnarray}

\par
Eigenvalues of the mass matrices are not the physical masses but the parameters in the Lagrangian. Theses quark masses (eigenvalues) are running masses which should be all taken on a single energy scale. In order to estimate the parameters $(\delta^q_1,\ \delta^q_2,\ \delta^q_3)$, we use the quark mass ratios. These mass ratios are, to a good approximation, independent of the energy scale, then the scale can be arbitrarily chosen \cite{GASSER}.  

\par 
For the values of the light and medium heavy quark $u,\ d,\ s$ and $c$ masses, we use the world average cited in Ref. \cite{PARTI}; 
\begin{eqnarray}
&& m_u = 2 - 8\ {\rm MeV}, \ \ \ \ \ m_c = 1.0 - 1.6\ {\rm GeV}, \nonumber \\
&& m_d = 5 - 15\ {\rm MeV}, \ \ \ \ \ m_s = 100-300\ {\rm MeV}, 
\end{eqnarray}
because these values are almost similar to the mass values at the scale $\mu=1{\rm GeV}$ \cite{GASSER}; $m_u=5.1\pm1.5{\rm MeV},\ m_d=8.9\pm2.6{\rm MeV},\ m_s=175\pm55{\rm MeV},\ m_c=1.35\pm0.05{\rm GeV}$.

\par 
For the heavy quark $b$ and $t$ masses, we estimate the running mass $m_q(\mu=1{\rm GeV})$ related to the physical mass $m_q^{phys}$ in the first order QCD as
\begin{equation}
m_q^{phys}=m_q(\mu=m_q)\left[1+\frac4{3\pi}\alpha_s(\mu=m_q)\right],
\end{equation}
where the running coupling constant $\alpha_s(\mu)$ and the running mass $m_q(\mu)$  are expressed as 
\begin{eqnarray}
&& \alpha_s(\mu)=\frac{g(\mu)^2}{4\pi}=\frac{4\pi}{\beta_0 L}\left(1-\frac{\beta_1}{\beta_0^2}\frac{{\rm ln}L}{L}\right), \nonumber \\
&& m_q(\mu)=\bar{m}\left(1-\frac{2\beta_1\gamma_0}{\beta_0^3}\frac{{\rm ln}L+1}{L}+\frac{8\gamma_1}{\beta_0^2}\frac1L\right)\left(\frac{L}2\right)^{-\frac{2\gamma_0}{\beta_0}}, \\
&& \ \ \ \ \beta_0=11-\frac23N_f,\ \ \gamma_0=2,\ \ \beta_1=102-\frac{38}3N_f,\nonumber \\
&& \ \ \ \ \gamma_1=\frac{102}{12}-\frac5{18}N_f, \ \ L={\rm ln}\frac{\mu^2}{\Lambda^2}. \nonumber
\end{eqnarray}
The estimated $m_b(\mu=1{\rm GeV})$ and $m_t(\mu=1{\rm GeV})$ from the physical mass $m_b^{phys}=4.3\pm0.2~{\rm GeV}$ and $m_t^{phys}=174\pm\at{22}{23}~{\rm GeV}$ \cite{PARTI} are 
\begin{eqnarray}
& \ \ \Lambda=0.1{\rm GeV} & \ \ \ \Lambda=0.2{\rm GeV} \nonumber \\
m_b(\mu=1{\rm GeV})= & 5.08\ \pm\ 0.28\ {\rm GeV}, & 5.49\ \pm\ 0.29\ {\rm GeV},\nonumber \\
m_t(\mu=1{\rm GeV})= & 289\ \pm\ 41\ {\rm GeV}, \ \ \ & 327\ \pm\ 48\ {\rm GeV},
\end{eqnarray}
for the flavor number $N_f=3$. We write the values of $m_b(\mu=1{\rm GeV})$ and $m_t(\mu=1{\rm GeV})$ for the renormalization group invariant scale $\Lambda=0.1{\rm GeV}$ and $\Lambda=0.2{\rm GeV}$ cases, because these mass values are sensitive to the values of $\Lambda$. Hereafter, we write the mass $m_q(\mu=1{\rm GeV})$ as $m_q$. From these mass values, we get the quark mass ratios, 
\begin{eqnarray}
&&\frac{m_u}{m_c}=0.0038 \pm 0.0025,\ \ \ \ \ \frac{m_d}{m_s}=0.050 \pm 0.035, \nonumber \\
&&\frac{m_c}{m_t}=0.0042 \pm 0.0013,\ \ \ \ \ \frac{m_s}{m_b}=0.038 \pm 0.019,
\end{eqnarray}
where we used the average values of $m_b(\mu=1{\rm GeV})$ and $m_t(\mu=1{\rm GeV})$ for $\Lambda=0.1{\rm GeV}$ and $\Lambda=0.2{\rm GeV}$ and involved the deviation from the average value in errors. 
\par
We estimated numerically the allowed regions of $(\delta^q_1,\ \delta^q_2,\ \delta^q_3)$ satisfying the constraint in which the ratios of the eigenvalues $(m_u,\ m_c,\ m_t)$ and $(m_d,\ m_s,\ m_b)$ of the mass matrices (5) are included in the experimental ranges of quark mass ratios (12). We showed the allowed regions for $(m_u,\ m_c,\ m_t)$ sector in Figs.~1(a), (b) ,(c) and for $(m_d,\ m_s,\ m_b)$ sector in Figs.~1(d), (e), (f). 

\par
\begin{center}
\setlength{\unitlength}{1cm}
\begin{picture}(7,1)
\put(0.5,0.5){\line(1,0){6}}
\end{picture}\\
Fig. 1 (a), (b), (c), (d), (e), (f)\\
\vspace{0.4cm}
\begin{picture}(7,1)
\put(0.5,0.5){\line(1,0){6}}
\end{picture}
\end{center}

The Figs.~1(a), (b), (c) represent the allowed regions of $(\delta^u_2,\delta^u_3 )$ plane corresponding to the $\delta^u_1=0.00005,\ 0.0001,\ 0.0004$ for $(u,\ c,\ t)$sector, respectively and Figs.~1(d), (e), (f) the allowed regions of  $(\delta^d_2,\delta^d_3 )$ plane corresponding to the $\delta^d_1=0.005,\ 0.01,\ 0.02$ for $(d,\ s,\ b)$ sector, respectively. For $\delta^u_1 < 0.000012$ and  $\delta^d_1 < 0.00085$, the allowed regions for $(\delta^u_2,\delta^u_3 )$ and $(\delta^d_2,\delta^d_3 )$ plane do not exist, respectively. It is seen in this Fig.~1 that the allowed regions for $(\delta^q_2, \delta^q_3)$ are symmetric with respect to the interchange between $\delta^q_2$ and $\delta^q_3$ . This symmetry is found easily in the approximate expressions for the eigenvalues and eigenvectors of the mass matrix (5). The eigenvalues are 
\begin{eqnarray}
&& m^q_1 \approx \left[\frac13(\delta^q_1+\delta^q_2+\delta^q_3)-\frac13\xi^q\right]\Gamma^q, \nonumber \\
&& m^q_2 \approx \left[\frac13(\delta^q_1+\delta^q_2+\delta^q_3)+\frac13\xi^q\right]\Gamma^q, \nonumber \\
&& m^q_3 \approx \left[3-\frac23(\delta^q_1+\delta^q_2+\delta^q_3)\right]\Gamma^q, 
\end{eqnarray}
where
\begin{equation}
\xi^q=\left[(2\delta^q_1-\delta^q_2-\delta^q_3)^2+3(\delta^q_2-\delta^q_3)^2\right]^{1/2}, 
\end{equation}
and the corresponding eigenvectors are $U^q_1$, $U^q_2$, $U^q_3$; 
\begin{eqnarray}
&& T^{\dagger}(\delta^q_1,\delta^q_2,\delta^q_3)=[(U^q_1),(U^q_2),(U^q_3)],\nonumber \\
&& T(\delta^q_1,\delta^q_2,\delta^q_3) \approx \left( 
   \begin{array}{ccc}
     \cos{\theta^q} & \sin{\theta^q} & \lambda^q\cos{\theta^q}+\mu^q\sin{\theta^q}    \\
     -\sin{\theta^q} & \cos{\theta^q} & -\lambda^q\sin{\theta^q}+\mu^q\cos{\theta^q}   \\  
   -\lambda^q & -\mu^q & 1 
   \end{array} \right)T_0, 
\end{eqnarray}
where
\begin{eqnarray}
&& \lambda^q=\frac1{3\sqrt{6}}(\delta^q_2-\delta^q_3), \ \ \ \ \mu^q=\frac1{9\sqrt{2}}(2\delta^q_1-\delta^q_2-\delta^q_3), \nonumber \\
&& \theta^q=\frac12\tan^{-1}\frac{\sqrt{3}(\delta^q_2-\delta^q_3)}{2\delta^q_1-\delta^q_2-\delta^q_3}.
\end{eqnarray}
Though these expressions are obtained approximately, the allowed regions for $(\delta^q_2,\delta^q_3)$ obtained from these approximate expressions are almost same as those in Fig.~1.

\par
Next we consider the CKM matrix $V$, 
\begin{equation}
V=T(\delta^u_1,\delta^u_2,\delta^u_3)T^{\dagger}(\delta^d_1,\delta^d_2,\delta^d_3).
\end{equation}
The matrix elements of $V$ are determined by various experiments, for example, nuclear beta decays, $K_{e3}$ decays, neutrino and antineutrino production of charm off valence $d$ quarks, neutrino production of charm, semileptonic decays of $B$ mesons produced on the $\Upsilon(4S)\ b\bar{b}$ resonance and {\it etc.} The absolute values for these matrix elements are tabulated as \cite{PARTI} 
\begin{equation}
V^{exp}=\left(
   \begin{array}{ccc}
   0.9747-0.9759 & 0.218-0.224 & 0.002-0.005 \\
   0.218-0.224 & 0.9738-0.9752 & 0.032-0.048 \\
   0.004-0.015 & 0.030-0.048 & 0.9988-0.9995
   \end{array}\right).
\end{equation}
We calculated  numerically the allowed regions of $(\delta^q_1, \delta^q_2, \delta^q_3)$ satisfying the restriction in which the absolute values of matrix elements of $V$ in Eq.~(17) are included in the experimental range of matrix elements (Eq.~(18)). First, we estimated the allowed regions of $(\delta^q_1, \delta^q_2, \delta^q_3)$ independent of the experimental constraint of ranges of the mass ratio (12). We showed the allowed region of $(\delta^u_2,\ \delta^u_3)$ for $(u,\ c,\ t)$ sector in Fig. 2(a) and of $(\delta^d_2,\ \delta^d_3)$ for $(d,\ s,\ b)$ sector in Fig.~2(b) fixing the $\delta^u_1$ and $\delta^d_1$ as $\delta^u_1=0.0001$ and $\delta^d_1=0.01$. For other values of $\delta^u_1$ and $\delta^d_1$, $\delta^u_1=0.00005,\ 0.0004$ and $\delta^d_1=0.005,\ 0.02$, the allowed regions for $(\delta^u_2,\ \delta^u_3)$ and $(\delta^d_2,\ \delta^d_3)$ are almost similar to those of the case $\delta^u_1=0.0001$ and $\delta^d_1=0.01$ shown in Fig .2. All combinations of all points in allowed region of $(\delta^u_2,\ \delta^u_3)$ with all points in allowed regions of $(\delta^d_2,\ \delta^d_3)$ are not allowed but the restricted combinations between some points in allowed region of $(\delta^u_2,\ \delta^u_3)$ and some points in allowed region of $(\delta^d_2,\ \delta^d_3)$ are allowed. For example, the combinations of a point in $(\delta^u_2,\ \delta^u_3)$ plain (shown by a large dot) with points in the area shown by the large dotts in $(\delta^d_2,\ \delta^d_3)$ plane are allowed. In Fig.~2, though we showed the solution corresponding to the case, $\delta_2^u > \delta_3^u$ and  $\delta_2^d < \delta_3^d$, there exist also the solutions corresponding to the cases,  $\delta_2^u < \delta_3^u$ and  $\delta_2^d > \delta_3^d$.  

\par
\begin{center}
\setlength{\unitlength}{1cm}
\begin{picture}(7,1)
\put(0.5,0.5){\line(1,0){6}}
\end{picture}\\
Fig. 2 (a), (b)\\
\vspace{0.4cm}
\begin{picture}(7,1)
\put(0.5,0.5){\line(1,0){6}}
\end{picture}
\end{center}

\par
As shown in Fig. 1 and Fig. 2, there is no common regions satisfying both constraints of the mass ratios and the CKM matrix in $(\delta^d_2,\ \delta^d_3)$ plane of $(d, \ s, \ b)$ sector. This fact is easily understood from the analytic expressions for $V$. Using the approximate expression (15) for $T(\delta_1^q,\delta_2^q,\delta_3^q)$, we can get the approximate expression for $V$ 
\begin{eqnarray}
 V \approx && \left(
  \begin{array}{cc}
   \cos(\theta^u-\theta^d) & \sin(\theta^u-\theta^d) \\
   -\sin(\theta^u-\theta^d) & \cos(\theta^u-\theta^d)\\
   -(\lambda^u-\lambda^d)\cos{\theta^d}-(\mu^u-\mu^d)\sin{\theta^d} & (\lambda^u-\lambda^d)\sin{\theta^d}-(\mu^u-\mu^d)\cos{\theta^d}
  \end{array} 
  \right. \nonumber \\
  &&\ \ \ \ \ \ \ \ \ \ \ \ \ \  \left.
  \begin{array}{c}
   (\lambda^u-\lambda^d)\cos{\theta^u}+(\mu^u-\mu^d)\sin{\theta^u} \\
    -(\lambda^u-\lambda^d)\sin{\theta^u}+(\mu^u-\mu^d)\cos{\theta^u} \\
    1
   \end{array} \right).
\end{eqnarray}
From this expression, for $\delta^d_2\approx\delta^d_3\gg\delta^d_1$, we can get the ratio $|V_{cb}/V_{tb}|\approx| -(\lambda^u-\lambda^d)\sin{\theta^u}+(\mu^u-\mu^d)\cos{\theta^u}|\approx|\mu^d|$ and then $\delta^d_2\approx\delta^d_3 > 0.20$ from the experimental ratio $|V_{cb}^{exp}/V_{tb}^{exp}| > 0.032$. On the other hand, from the ratio of mass eigenvalues (13), for $\delta^d_2\approx\delta^d_3\gg\delta^d_1$, we can get $m^d_2/m^d_3 \approx 2(\delta^d_2+\delta^d_3)/9$ then $\delta^d_2\approx\delta^d_3 < 0.13$ from the experimental range $m_s/m_b=0.038\pm0.019$. We comment here on the difference between the values calculated by Eq.~(19) and by numerically exact procedure. The values for $V_{us}$ and $V_{cd}$ elements calculated by Eq.~(19) are different from the values calculated by numerical and exact procedure about 2\%, and for $V_{ub}$, $V_{cb}$, $V_{td}$ and $V_{ts}$ about 20\% and for $V_{ud}$, $V_{cs}$ and $V_{tb}$ about 0.1\%.   

\par
Because of the fact that there is no common region in $(\delta^d_2,\ \delta^d_3)$ plane satisfying the quark mass ratios and the CKM matrix, we consider the case containing the phases $\varphi^q_i$ in the quark mass matrix as Eq.~(3). Although there are 6  degrees of freedom for phases $\varphi^q_i$, only two phases  $\varphi^d_2$ and $\varphi^d_3$ are considered in our analysis because  only $\delta^d_2$ and $\delta^d_3$ in the violation parameters are about 0.1 and other parameters are extremely small ($\delta^d_1\sim0.01$, $\delta^u_1\sim0.0001$, $\delta^u_{2,3}\sim0.01$), then the phases with these other parameters scarcely contribute to CKM matrix in contrast to two phases $\varphi^d_2$ and $\varphi^d_3$. We parametrize the $(d,s,b)$ sector quark mass matrix using the very small phases $\phi_2$ and $\phi_3$ instead of the phases $\varphi_2^d$ and $\varphi_3^d$, as 
\begin{equation}
M^d=\Gamma^d\left(
 \begin{array}{ccc}
   1 & 1-\delta^d_1 & (1-\delta^d_2)e^{i\phi_2} \\
   1-\delta^d_1 & 1 & (1-\delta^d_3)e^{i\phi_3} \\
   (1-\delta^d_2)e^{-i\phi_2} & (1-\delta^d_3)e^{-i\phi_3} & 1
  \end{array}\right), \ \ \ \ \phi_i \ll 1, \ \ \ \ (i=2,\ 3)
\end{equation}
and for $M^u$ we use the type of Eq.~(5) with no phase. The approximate expressions for the eigenvalues of the mass matrix (20) are the same expressions for $m^d_1,\ m^d_2$ and $m^d_3$ as Eq.~(13), but the expression for $\xi^d$ is changed to containing the phases $\phi_2$ and $\phi_3$ as follows,  
\begin{equation}
\xi^d=\left[(2\delta^d_1-\delta^d_2-\delta^d_3)^2+3(\delta^d_2-\delta^d_3)^2+3(\phi_2-\phi_3)^2\right]^{1/2}. 
\end{equation}
The expression for CKM matrix is given in this approximation as     
\begin{eqnarray}
 V & = & T(\delta^u_1,\delta^u_2,\delta^u_3)T^{\dagger}(\delta^d_1,\delta^d_2,\delta^d_3,\phi_2,\phi_3) \nonumber \\
 & \approx & \left(
  \begin{array}{cc}
   \cos{\theta^u}c^d+\sin{\theta^u}s^{d\last} & -\cos{\theta^u}s^d+\sin{\theta^u}c^d  \\
   -\sin{\theta^u}c^d+\cos{\theta^u}s^{d\last} & \sin{\theta^u}s^d+\cos{\theta^u}c^d  \\
   -(\lambda^u-\lambda^{d\last})c^d-(\mu^u-\mu^{d\last})s^{d\last} & (\lambda^u-\lambda^{d\last})s^d-(\mu^u-\mu^{d\last})c^d
   \end{array} 
  \right. \nonumber \\
  &&\ \ \ \ \ \ \ \ \ \ \ \ \ \  \left.
  \begin{array}{c}
   (\lambda^u-\lambda^d)\cos{\theta^u}+(\mu^u-\mu^d)\sin{\theta^u} \\
    -(\lambda^u-\lambda^d)\sin{\theta^u}+(\mu^u-\mu^d)\cos{\theta^u} \\
    1
   \end{array} \right),
\end{eqnarray}
where 
\begin{eqnarray}
&& c^d=\frac{\sqrt{\xi^d-(2\delta^d_1-\delta^d_2-\delta^d_3)}}{\sqrt{2\xi^d}}, \nonumber \\
&& s^d=\frac{-\sqrt{3}\left\{(\delta^d_2-\delta^d_3)-i(\phi_2-\phi_3)\right\}}{\sqrt{2\xi^d}\sqrt{\xi^d-(2\delta^d_1-\delta^d_2-\delta^d_3)}}, \nonumber \\
&& \lambda^d=\frac1{3\sqrt{6}}\left\{(\delta^d_2-\delta^d_3)-i(\phi_2-\phi_3)\right\}, \nonumber \\
&& \mu^d=\frac1{9\sqrt{2}}\left\{2\delta^d_1-(\delta^d_2+\delta^d_3)-3i(\phi_2+\phi_3)\right\}, 
\end{eqnarray}
and $\theta^u$, $\lambda^u$ and $\mu^u$ are given in Eq.~(16). We calculated  the allowed regions of $(\delta_1^u,\ \delta_2^u,\ \delta_3^u)$ and  $(\delta_1^d,\ \delta_2^d,\ \delta_3^d,\ \phi_2,\ \phi_3)$ numerically and exactly satisfying two constraints of the experimental values of ranges of the quark mass ratios Eq.~(12) and the CKM matrix $V^{exp}$ Eq.~(18) and showed these regions in Fig.~3. We showed the case $(\delta_1^u,\ \delta_1^d,\ \phi_2,\ \phi_3)$=(0.00005,\ 0.005,\ $-4^{\circ}$,\ $-4^{\circ})$ in Fig.~3(a), (0.0001,\ 0.01,\ $-4^{\circ}$,\ $-4^{\circ}$) in Fig.~3(b). Solutions corresponding to the cases other than $(\phi_2,\ \phi_3)\approx(-4^{\circ},\ -3^{\circ} \sim -4^{\circ})$ do not exist. 

\par
\begin{center}
\setlength{\unitlength}{1cm}
\begin{picture}(7,1)
\put(0.5,0.5){\line(1,0){6}}
\end{picture}\\
Fig. 3 (a), (b)\\
\vspace{0.4cm}
\begin{picture}(7,1)
\put(0.5,0.5){\line(1,0){6}}
\end{picture}
\end{center}

\par 
In order to see the effects of the CP violation, we rephase the CKM matrix $V=T(\delta^u_1,\delta^u_2,\delta^u_3)T^{\dagger}(\delta^d_1,\delta^d_2,\delta^d_3,\phi_2,\phi_3)$ to the standard parametrized CKM matrix $V^R$ where the matrix elements $V^R_{ud}$, $V^R_{us}$, $V^R_{cb}$ and $V^R_{tb}$ are real number, by using the rephasing matrix $P_u$ and $P_d$ as 
\begin{eqnarray}
&&V^{R} = P_uVP^{\dagger}_d, \nonumber \\
&&P_u={\rm diag}[e^{i\alpha'},\ 1,\ e^{i\beta'}],\ \ \  P_d={\rm diag}[e^{i\alpha},\ 1,\ e^{i\beta}].  
\end{eqnarray}
The parameters $\alpha$, $\alpha'$, $\beta$ and $\beta'$ are determined as 
\begin{eqnarray}
&& \alpha=\tan^{-1}\frac{{\rm Im}V_{ud}}{{\rm Re}V_{ud}}-\tan^{-1}\frac{{\rm Im}V_{us}}{{\rm Re}V_{us}},\ \ \ \ \ \ \ \  \alpha'=-\tan^{-1}\frac{{\rm Im}V_{us}}{{\rm Re}V_{us}}, \nonumber \\
&& \beta=\tan^{-1}\frac{{\rm Im}V_{cb}}{{\rm Re}V_{cb}},\ \ \ \ \ \ \ \  \beta'=\tan^{-1}\frac{{\rm Im}V_{cb}}{{\rm Re}V_{cb}}-\tan^{-1}\frac{{\rm Im}V_{tb}}{{\rm Re}V_{tb}}. 
\end{eqnarray}
In the standard parametrized CKM matrix $V^R$ in which the element $V^R_{cd}$ is almost real as recognized in the Wolfenstein parametrization \cite{WOLFEN}, the parameters $\rho$ and $\eta$ characterizing the CP violation which are the vertex coordinate of unitarity triangle are expressed as
\begin{equation}
\rho=\frac{{\rm Re}(V^{R\last}_{ub}V^R_{ud})}{|V^{R\last}_{cb}{V^R_{cd}}|},\ \ \ \ \ \ \ \ \ \ \eta=-\frac{{\rm Im}(V^{R\last}_{ub}V^R_{ud})}{|V^{R\last}_{cb}{V^R_{cd}}|}.
\end{equation}
The phenomenological constraints for parameters $\rho$ and $\eta$ has been examined by Pich and Prades \cite{PICH} using the recent information on the non-perturbative hadronic inputs needed in the analysis of $B^0-\bar{B}^0$ mixing and the CP-violating parameter $\varepsilon$ of $K^0-\bar{K}^0$ system. They gave the results of parameters $\rho$ and $\eta$ for the best estimate set of input parameters. We showed our results of $(\rho,\ \eta)$ for $(\delta_1^u,\ \delta_1^d,\ \phi_2,\ \phi_3)$ fixed as $(0.00005,\ 0.005,\ -4^{\circ},\ -4^{\circ})$ and $(0.0001,\ 0.01,\ -4^{\circ},\ -4^{\circ})$ in Fig. 4, besides the Pich and Prades results which are surround by circles centered at $(0, 0)$ and $(1, 0)$ and hyperbola correspond to the input parameters ($\chi_d=0.76\pm0.06$, $m_t^{pole}=174\pm16~{\rm GeV}$, $\tau(B^0_d)=1.61\pm0.08~{\rm ps}$, $\tau(B^0_d)|V_{cb}|^2=(3.9\pm0.6)\times10^9{\rm GeV}^{-1}$, $|V_{ub}/V_{cb}|=0.08\pm0.03$, $\widehat{B}_K=0.50\pm0.15$, $\hat{\xi}_B/f_{\pi}=2.0\pm0.5$) \cite{PICH}. From this Fig. 4, we can say that the values of parameters $(\delta_1^u,\ \delta_2^u,\ \delta_3^u)$ and $(\delta_1^d,\ \delta_2^d,\ \delta_3^d,\ \phi_2,\ \phi_3)$ in allowed regions shown in Fig. 3 are almost consistent with the phenomenological CP violation results. 

\par
\begin{center}
\setlength{\unitlength}{1cm}
\begin{picture}(7,1)
\put(0.5,0.5){\line(1,0){6}}
\end{picture}\\
Fig. 4\\
\vspace{0.4cm}
\begin{picture}(7,1)
\put(0.5,0.5){\line(1,0){6}}
\end{picture}
\end{center}

\par
We summarized the ranges of parameters $(\delta_1^u,\ \delta_+^u\equiv(\delta_2^u+\delta_3^u)/2,\ \delta_-^u\equiv \delta_2^u-\delta_3^u)$ and $(\delta_1^d,\ \delta_+^d\equiv(\delta_2^d+\delta_3^d)/2,\ \delta_-^d\equiv \delta_2^d-\delta_3^d,\ \phi_+\equiv(\phi_2+\phi_3)/2,\ \phi_-\equiv \phi_2-\phi_3)$ obtained in previous analysis in Table 1. The compound signs with the values of $\delta_-^u$, $\delta_-^d$ and $\phi_-$ correspond to each other. Of course, all combinations of values in the ranges shown in Table 1 are not the solutions but special combinations are the solutions.

\begin{table}
\begin{center}
\caption[Table 1]{The ranges of parameters $(\delta_1^u,\ \delta_+^u,\ \delta_-^u,\ \delta_1^d,\ \delta_+^d,\ \delta_-^d,\ \phi_+,\ \phi_-)$ satisfying the mass ratios (Eq.~(12)) and CKM matrix (Eq.~(18)).}
\label{Table}
\vspace{0.5cm}
 \begin{tabular}{|l|c|l|c|}   \hline
  \multicolumn{2}{|c|}{uct sector} & \multicolumn{2}{|c|}{dsb sector} \\ \hline
  $\ \delta_1^u\ $ &\ $0.00001 \sim 0.0004$\ & $\ \delta_1^d\ $ & $\ 0.001 \sim 0.015$\ \\  
  $\ \delta_+^u\equiv(\delta_2^u+\delta_3^u)/2\ $ &\ $0.0064 \sim 0.0125$\ & $\ \delta_+^d\equiv(\delta_2^d+\delta_3^d)/2\ $ & $\ 0.040 \sim 0.129$\ \\
  $\ \delta_-^u\equiv \delta_2^u-\delta_3^u\ $ &\ $\pm(0.0 \sim 0.0043)$\ & $\ \delta_-^d\equiv \delta_2^d-\delta_3^d\ $ & $\ \pm(-0.038 \sim -0.006)$\  \\  
  $                  $ &                          & $\ \phi_+^d\equiv(\phi_2+\phi_3)/2\ $ &\ $-4^{\circ} \sim -3^{\circ}$\ \\  
  $                  $ &                          & $\ \phi_-^d\equiv \phi_2-\phi_3\ $ &\ $\pm(-1^{\circ} \sim 0^{\circ})$\ \\ \hline
 \end{tabular}
\end{center}
\end{table}%

We show the typical solutions, and the mass ratios, CKM matrix elements and $(\rho,\ \eta)$ corresponding to these solutions;
\begin{eqnarray}
 {\rm solution\ A}: && \left\{ \begin{array}{ll}
 \delta_1^u=0.00005,\ \ \delta_2^u=0.01,\ \ \delta_3^u=0.009, \\
 \delta_1^d=0.005,\ \ \delta_2^d=0.054,\ \ \delta_3^d=0.08,\ \ \phi_2=-4^{\circ},\ \ \phi_3=-4^{\circ}, 
 \end{array} \right. \nonumber \\
 && \frac{m_u}{m_c}=0.0019,\ \frac{m_c}{m_t}=0.0042,\ \frac{m_d}{m_s}=0.025,\ \frac{m_s}{m_b}=0.031,\ \nonumber \\
&& V=\left(
   \begin{array}{ccc}
   0.9755 & 0.2198 & 0.0036 \\
   0.2196 & 0.9750 & 0.0340 \\
   0.0077 & 0.0333 & 0.9994
   \end{array}\right), \nonumber \\
&& \rho=0.088,\ \ \eta=0.47, \nonumber \\   
 {\rm solution\ B}: && \left\{ \begin{array}{ll}
 \delta_1^u=0.0001,\ \ \delta_2^u=0.01,\ \ \delta_3^u=0.009, \\
 \delta_1^d=0.01,\ \ \delta_2^d=0.07,\ \ \delta_3^d=0.102,\ \ \phi_2=-4^{\circ},\ \ \phi_3=-4^{\circ},
 \end{array} \right. \nonumber \\
 && \frac{m_u}{m_c}=0.0058,\ \ \frac{m_c}{m_t}=0.0042,\ \ \frac{m_d}{m_s}=0.058,\ \ \frac{m_s}{m_b}=0.040,\ \nonumber \\
&& V=\left(
   \begin{array}{ccc}
   0.9753 & 0.2210 & 0.0043 \\
   0.2210 & 0.9747 & 0.0347 \\
   0.0087 & 0.0339 & 0.9994
   \end{array}\right), \nonumber \\   
&& \rho=0.022, \ \ \eta=0.56.     
\end{eqnarray} 

\vspace{1.2cm}

\begin{large}
\par \noindent
{\bf 3. Discussions}
\end{large}
\vspace{0.5cm}
\par
We analysed precisely the mass hierarchy of quarks and CKM matrix in the law energy by the universal Yukawa coupling model with small violations (Eq.(3)). Violation parameters estimated are 8: ($\delta_1^u$,\ $\delta_2^u$,\ $\delta_3^u$,\ $\delta_1^d$,\ $\delta_2^d$,\ $\delta_3^d$,\ $\phi_2$,\ $\phi_3$), and estimated values of these are tabulated in Table 1. Other phases than $\phi_2$ and $\phi_3$ do not contribute to our present analysis because of extreme smallness of ($\delta_1^u$,\ $\delta_2^u$,\ $\delta_3^u$,\ $\delta_1^d$). We fitted 8 violation parameters to 8 experimental data:  4 quark mass ratios (Eq.~(12)), 3 mixing angles determined by CKM matrix elements (Eq.~(18)) and 1 phase determined by CP violation which relates to $\rho$ and $\eta$. 
 
\par 
Here we comment on the differences between our model and others \cite{FRITZSCH}-\cite{HABA}. The mass matrices depending on models adopted and CKM matrix are connected through the unitary matrices $T_{u, d}$ depending on models as follows; 
\begin{eqnarray}
&&T_uM^uT_u^{-1}=M^u_D={\rm diag}[m_u,\ m_c,\ m_t], \nonumber \\
&&T_dM^dT_d^{-1}=M^d_D={\rm diag}[m_d,\ m_s,\ m_b], \nonumber \\
&&V=T_uT^{\dagger}_d.
\end{eqnarray}
From $V=T_uT^{\dagger}_d$ and $V^{exp} \approx 1$,  Fritzsch type , Stech type model and many other models \cite{FRITZSCH}, \cite{HABA} adopt the unitary matrices $T_{u,d}^{\rm Fritzsch\ type}$ as $T_{u,d}^{\rm Fritzsch\ type} \approx 1$ and then $(M^u)^{\rm Fritzsch\ type}=(T_u^{\rm Fritzsch\ type})^{-1}{\rm diag}[m_u,\ m_c,\ m_t]T_u^{\rm Fritzsch\ type}\approx{\rm diag}[m_u,\ m_c,\ m_t]$ and $(M^d)^{\rm Fritzsch\ type}=(T_d^{\rm Fritzsch\ type})^{-1}{\rm diag}[m_d,\ m_s,\ m_b]T_d^{\rm Fritzsch\ type}\approx{\rm diag}[m_d,\ m_s,\ m_b]$. However, if we change $T_u \to T_uS$ and $T_d \to T_dS$ where $S$ is some arbitrary unitary matrix, the CKM matrix $V$ remains unchanged. The democratic model and the universal Yukawa coupling model \cite{HARARI}, \cite{GATTO}, in fact, use the following uitary matrices $T_{u,\ d}^{\rm universal\ coupling\ type}$ as
\begin{equation}
T_{u,\ d}^{\rm universal\ coupling\ type} \approx T_{u,\ d}^{\rm Fritzsch\ type}T_0,
\end{equation}
where $T_0$ is the unitary matrix defined in Eq.~(2). CKM matrix does not depend on the unitary matrices $T_{u, d}$ adopted but the weak interaction eigenstates do on it then on the mass matrices adopted. 

\par
Even in Fritzsch type models, there are many parametrizations and all parametrizations cannot explain precisely the present quark mass hierarchy and CKM matrix. For example, original Fritzsch model can not explain the observed large top quark mass and modified model (see the literature of B.Dutta and S. Nandi in \cite{FRITZSCH}) can explain the large top quark mass but has to use the up quark mass matrix in which the $(2, 3)$ and the $(3,2)$ elements are unequal. Recent Peccei and Wang analysis (the literature in \cite{HABA}) uses the mass matrix $M^u=T_u^{-1}\{{\rm diag}~m_t[\xi_{ut}\lambda^7,\ \xi_{ct}\lambda^4,\ 1]\}T_u\approx{\rm diag}~m_t[\xi_{ut}\lambda^7,\ xi_{ct}\lambda^4,\ 1]$ and $M^d=T_d^{-1}\{{\rm diag}~m_b[\xi_{db}\lambda^4,\ \xi_{sb}\lambda^2,\ 1]\}T_d\approx{\rm diag}~m_b[\xi_{db}\lambda^4,\ \xi_{sb}\lambda^2,\ 1]$, where $\xi_{ut} = 0.49,\ \xi_{ct} = 1.46,\ \xi_{db} = 0.58,\ \xi_{sb} = 0.55$ and $\lambda$ is the parameter in Wolfenstein parametrization \cite{WOLFEN} of CKM matrix, and takes the mass matrices explaining the law energy data precisely. But the principle to take their mass matrices is not so clear. 
 
\par
We will analyse the problem of neutrino mixing using the present our model in next work. As we mentioned above, the weak interaction eigenstates depend on the unitary matrices $T_{u,d}$ adopted, then the analysis of mass matrix involving the lepton sector like the neutrino mixing problem will give the clue to check the validity of our model and other models.      

\newpage

\newpage
\par \noindent
\begin{center}
\begin{large}
Figuare captions 
\end{large}
\end{center}

\par \noindent
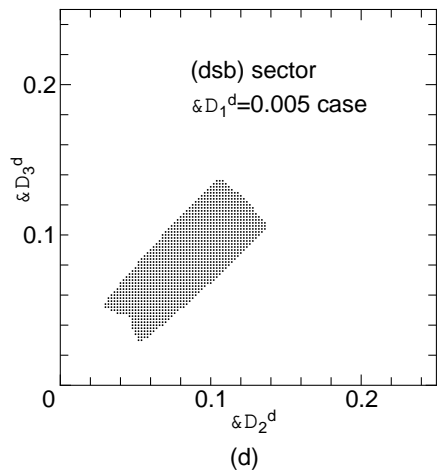
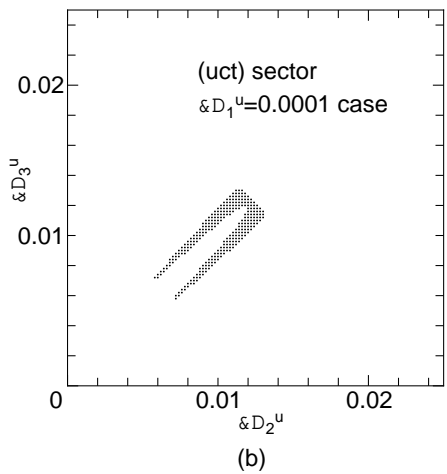
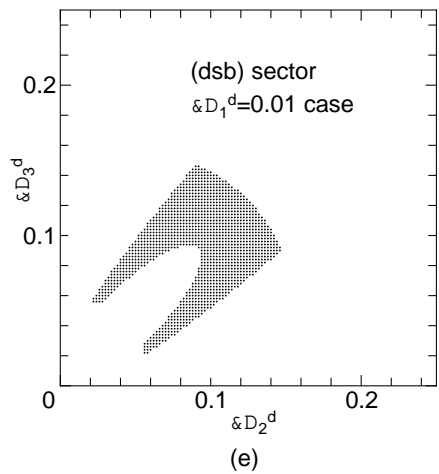
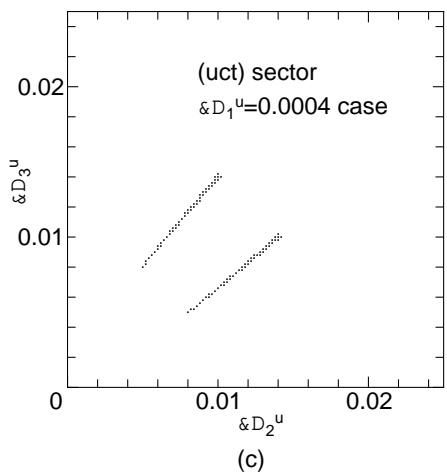
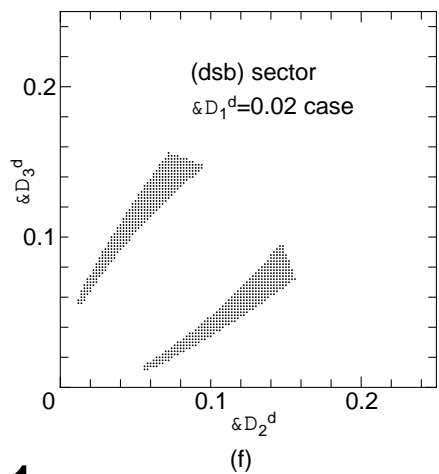
Fig.~1. The allowed regions for $(\delta_1^u,\ \delta_2^u,\ \delta_3^u)$ and $(\delta_1^d,\ \delta_2^d,\ \delta_3^d)$ satisfying the mass ratios (Eq.~(12)). (a), (b), (c): The allowed regions of $(\delta_2^u, \delta_3^u)$ plane corresponding to the $\delta_1^u=0.00005,\ 0.0001,\ 0.0004$  for $(u, c, t)$ sectors, respectively. (d), (e), (f): The allowed regions of $(\delta_2^d, \delta_3^d)$ plane corresponding to the $\delta_1^d=0.005,\ 0.01,\ 0.02$ for $(d, s, b)$ sectors, respectively.

\vspace{0.5cm}
\par \noindent
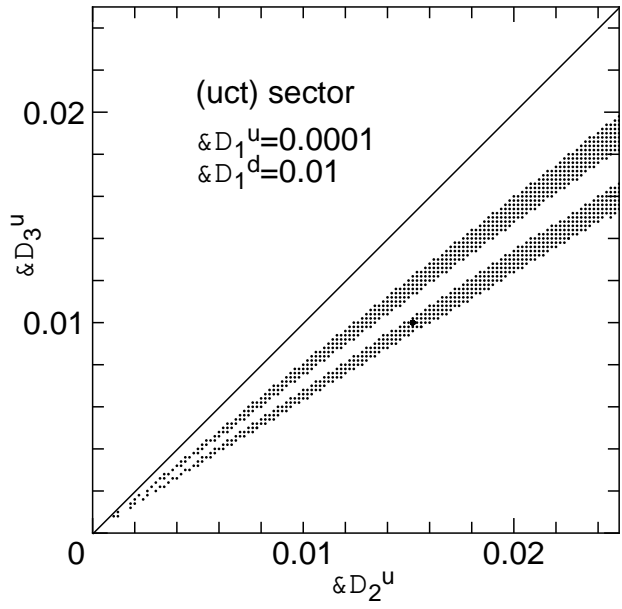
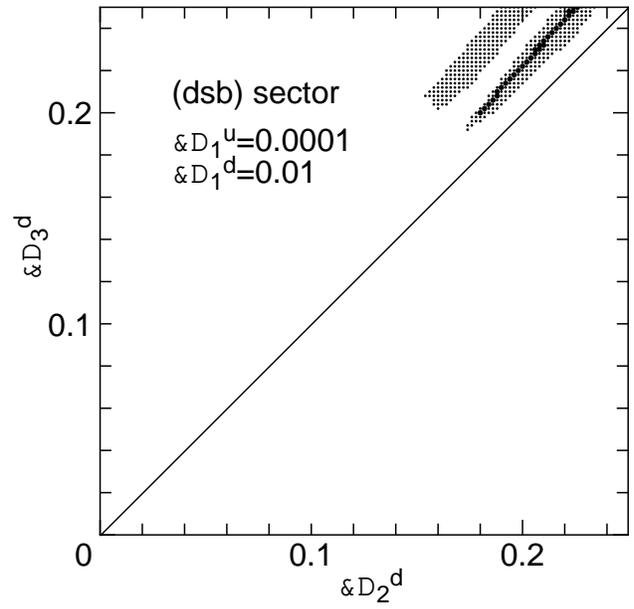
Fig.~2. The allowed regions of $(\delta_1^u,\ \delta_2^u,\ \delta_3^u)$ and $(\delta_1^d,\ \delta_2^d,\ \delta_3^d)$ satisfying the experimental CKM matrix elements (Eq.~(18)). (a): The allowed region of $(\delta_2^u, \delta_3^u)$ plane corresponding to the $\delta_1^u=0.0001$, $\delta_1^d=0.01$. (b): The allowed regions of $(\delta_2^d, \delta_3^d)$ plane corresponding to the $\delta_1^u=0.0001$, $\delta_1^d=0.01$.  All combinations of all points in allowed region of $(\delta^u_2,\ \delta^u_3)$ with all points in allowed regions of $(\delta^d_2,\ \delta^d_3)$ are not allowed. For example, the combinations of a point in $(\delta^u_2,\ \delta^u_3)$ plain (shown by a large dot) with points in the area shown by the large dotts in $(\delta^d_2,\ \delta^d_3)$ plane are allowed.

\vspace{0.5cm}
\par \noindent
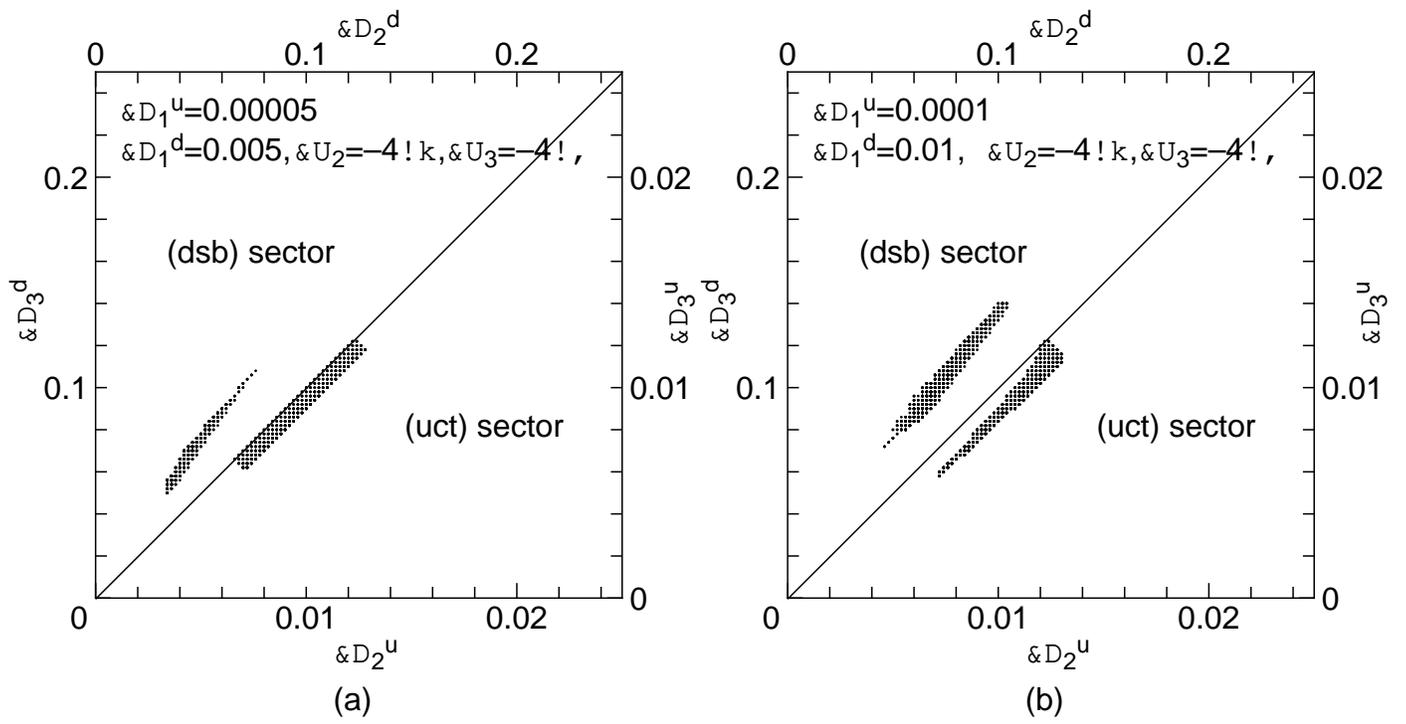
Fig.~3.  The allowed regions of $(\delta_1^u,\ \delta_2^u,\ \delta_3^u)$ and $(\delta_1^d,\ \delta_2^d,\ \delta_3^d)$ satisfying the mass ratios (Eq.~(12)) and the experimental CKM matrix elements (Eq.~(18)).  (a): the allowed regions of $(\delta_2^u, \delta_3^u)$ and $(\delta_2^u, \delta_3^u)$ planes corresponding to $\delta_1^u=0.00005$, $\delta_1^d=0.005$, $\phi_2=-4^{\circ}$, $\phi_3=-4^{\circ}$. (b): the allowed regions of $(\delta_2^u, \delta_3^u)$ and $(\delta_2^u, \delta_3^u)$ planes corresponding to $\delta_1^u=0.0001$, $\delta_1^d=0.01$, $\phi_2=-4^{\circ}$, $\phi_3=-4^{\circ}$.

\vspace{0.5cm}
\par \noindent
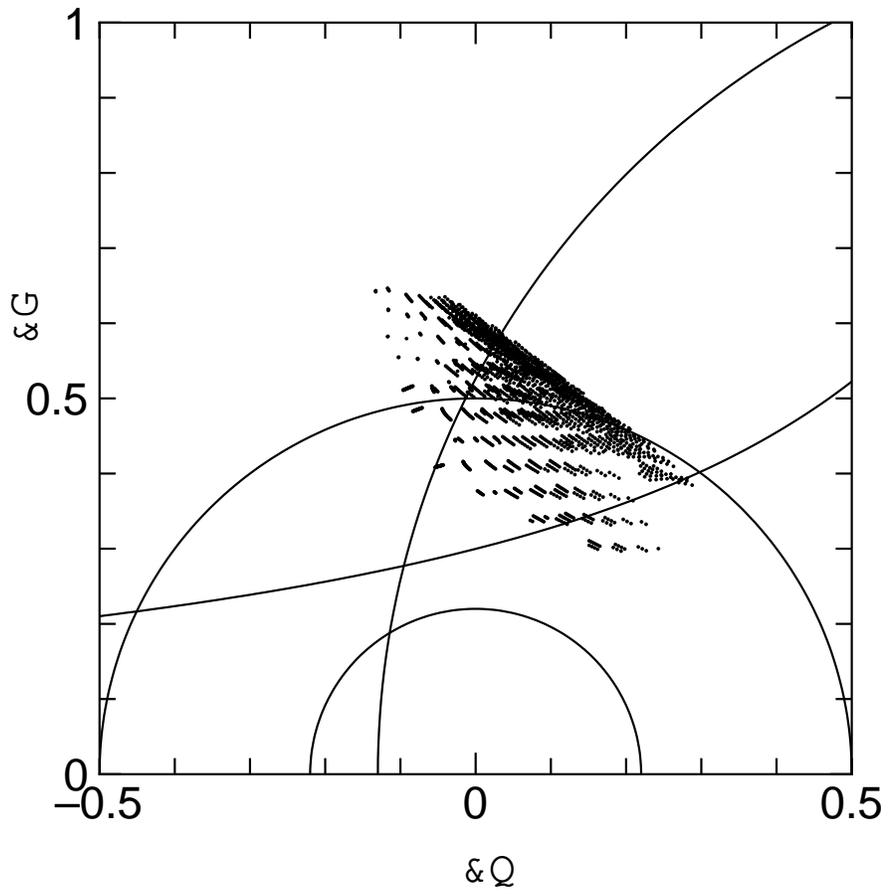
Fig.~4. The parameters ($\rho$,\ $\eta$) satisfying the mass ratios (Eq.~(12)) and the experimental CKM matrix elements (Eq.~(18)) for the $(\delta_1^u,\ \delta_1^d,\ \phi_2,\ \phi_3)$ fixed as $(0.00005,\ 0.005,\ -4^{\circ},\ -4^{\circ})$ and $(0.0001,\ 0.01,\ -4^{\circ},\ -4^{\circ})$. Area surrounded by the circles centered at $(0, 0)$ and $(1, 0)$ and hyperbola is the allowed region for CP violation given by Pich and Prades \cite{PICH}.

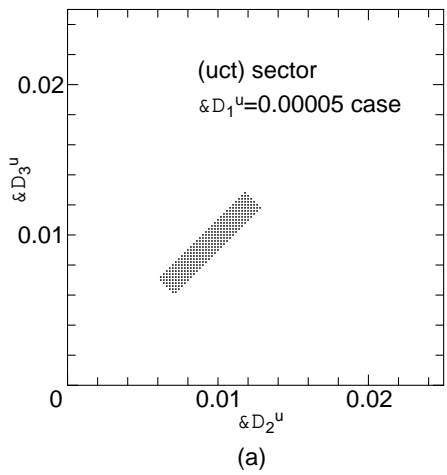
\end{document}